# Optical manipulation of intrinsic localized vibrational energy in cantilever arrays

M. Sato[1], B. E. Hubbard[1], A. J. Sievers[1], B. Ilic[2,3], and H. G. Craighead[3]

[1]Laboratory of Atomic and Solid State Physics
Cornell University, Ithaca, NY 14853-2501, USA
[2]Cornell Nanofabrication Facility
Cornell University, Ithaca, NY 14853-5403, USA
[3]Department of Applied and Engineering Physics
Cornell University, Ithaca, NY 14853-3501, USA

**Abstract**

Optically-induced real-time impurity modes are used to shepherd intrinsic localized vibrational modes (discrete breathers) along micromechanical arrays via either attractive or replulsive interactions. Adding an electrode to the cantilever array provides control of the sign of lattice anharmonicity, hence allowing both hard and soft nonlinearities to be studied. A number of dynamical effects are demonstrated and explained, including the optical tweezing of localized vibrational energy in a nonlinear lattice.

PACS. 05.45.-a    Nonlinear dynamics and nonlinear dynamical systems.

PACS. 63.20.Pw   Localized modes

PACS. 85.85.+j    Micro- and nano-electromechanical systems (MEMS/NEMS) and devices

A wide variety of periodic, nonlinear systems have been predicted to support spatially localized excitations in addition to plane-wave ones[1-7]. In order to better understand and visualize the general experimental properties of such nonlinear excitations different macroscopic lattice systems have been developed[8-12]. Particularly enlightening have been those studies which use laser observational methods. These range from the technique of low temperature laser scanning spectroscopy to observe localized excitations (discrete breathers) in Josephson junction arrays[13], to the optical imaging of localized vibrational modes in cantilever arrays[14], to the identification of discrete solitons in optically-induced real time waveguide arrays[15]. Although there is experimental evidence that intrinsic localized modes (ILM), also called discrete breathers (DB), can be generated in atomic lattices[16-19] the question remains how to manipulate such nanoscale excitations.

In this Letter it is demonstrated experimentally that optically-induced real-time impurity modes can be used to either repel or attract an ILM/DB in a nonlinear lattice. For a variety of di-element cantilever arrays with hard anharmonicity a laser-induced impurity mode is observed to repel an ILM/DB. By adding an electrode[20] to a di-element or mono-element cantilever array the anharmonicity can be converted to the soft type (like atomic lattices) and the ILM/DB is then attracted to the impurity mode. Our experiments suggest that as long as the size of the induced impurity mode is comparable to the size of the ILM/DB such processes can occur.

The experimental setup for the interaction study of a laser-induced impurity and an ILM/DB is shown in Fig. 1(a). The coupled cantilevers are made from a 300 nm thick $Si_3N_4$ film. The lengths of the longer and shorter cantilevers are 55 and 50 μm, while the width and pitch are 15 and 65 μm, respectively. A laser line illuminator coupled to a 1D-CCD is used to observe the excitation pattern in the cantilever array produced by a piezo-electric transducer (PZT) as described previously[14]. In order to create a movable impurity mode a fiber-coupled IR diode laser is used to generate a spot size of about 200 microns which heats a few cantilevers of the array, as shown in Fig. 1(b). Earlier studies with single micromechanical oscillators have shown that heating decreases its Young's modulus and hence its spring constant[21,22]. The spot from a low intensity HeNe laser is used only to measure the linear impurity mode spectrum induced by the high intensity IR laser. Since the nonlinearity of this cantilever array is positive [14] the interaction of a gap impurity mode with a hard ILM/DB can be studied.

To examine the physically relevant case of an ILM/DB in a lattice with soft anharmonicity, interacting with a soft linear-induced impurity mode, a gold film is deposited on the cantilevers. Beneath the cantilevers is a thermal oxide insulating layer so that a bias voltage can be applied between the gold film and the conductive Si substrate. Depending on the applied voltage the sign of the nonlinearity of a cantilever array can be changed from hard to soft[20]. Thus, soft nonlinear ILM/DB can be generated with the combination of the PZT driver and the DC bias voltage applied between the cantilevers and the substrate.

To demonstrate that the local IR (~100 mW) laser heating effect produces an impurity mode in the linear response spectrum of a di-element array of the hard anharmonicity type, a weak HeNe probe laser is focused a few lattice sites away from the IR laser spot. The di-element cantilever linear response spectra are then measured by monitoring the probe laser while sweeping the PZT frequency at small amplitude, both with the heating-laser on and off. The experimental mode spectra for the optic and acoustic bands for both cases are shown in Fig. 2. Below the narrow optic band there is an isolated feature (see arrow) in Fig. 2 (a) that indicates the formation of the gap linear impurity mode when the heating laser is on. Because of the much larger bandwidth of the lower branch the local mode is too close to the acoustic branch to be observed in Fig. 2(b). These findings are consistent with linear eigenfrequency calculations. From the comparison with the eigenfrequency calculation, the relative spring constant change for the shorter cantilever (center of the mode) is estimated to be -10%.

Both the repulsive and the attractive interactions between a laser induced impurity mode and an ILM/DB made from the corresponding positive or negative nonlinear band states are shown in Fig. 3. Figure 3 (a) consists of several frames taken with a 1D-CCD camera. The finely spaced horizontal lines are images of individual cantilevers. The dark region in each frame is a few ms long and shows a stationary ILM/DB. The arrow on the left identifies the ILM/DB starting point. The white rectangles identify the IR laser spot. The arrow on the left identifies the laser starting point. The laser power is typically 50-100 mW and the spot size is 100-200 μm.

The traces in Fig. 3(a) are made as follows. First, a single ILM/DB is created by using a chirped driver, which is then followed by a CW drive (constant amplitude and constant frequency) to produce a locked ILM/DB as described in Ref. [14]. Next, the heating IR laser is turned on and gradually moved towards to the ILM/DB. For the hard nonlinear array, case (a), at about frame 3 the ILM/DB moves away from the laser spot with a hopping motion. To observe this repulsion, the laser spot should be on the envelope of the ILM/DB. Once the ILM/DB hops, it is stable at this new position if the laser spot is moved far from the ILM/DB (see frames 10, 40 and 52). The ILM/DB can be moved to any location in the lattice using this method. Residual manufacturing impurity effects exist around frames 14 and 44 so that the ILM/DB hops a large distance. When the laser power is weak, it has no effect on the ILM/DB (see the frames near 24).

To study experimentally the gap impurity mode-ILM/DB interaction in a soft anharmonic lattice only a mono-element type array is required. Since the bottom of the dispersion curve is at $k = 0$ this mode can be excited by the PZT, and the large amplitude modulational instability used to produce ILM/DB. Experimental data demonstrating the attractive interaction between the impurity mode and the ILM/DB for the soft nonlinear array are shown in Fig. 3(b). Because of the large bandwidth of this coupled cantilever design a strong PZT excitation is required to separate off a locked ILM/DB. For this reason a background of running modes is conspicuous and, unlike Fig. 3(a), the horizontal lines of the remaining stationary cantilevers cannot be seen here. Since this ILM/DB envelope is oscillating about its equilibrium position, the pinning phenomenon is effectively weaker and the ILM/DB moves more easily than the one shown in Fig. 3(a). As the white laser spot approaches the ILM/DB in Fig. 3(b) the ILM/DB remains fixed until frame 5 where it is attracted and captured. We find that this happens even if the heating laser spot has not yet reached the envelope of the ILM. Similar phenomena are observed in simulations based on the model described below. Once the ILM/DB is captured, it follows the motion of the laser spot as shown in Fig. 3(b). Note that at frames 10, 17, and 21, the ILM/DB remains in place while the heating laser is turned off.

Using numerical simulations the general nature of these and other experimental results can be characterized. The equation of motion of a single cantilever is

$$m_i \frac{d^2}{dt^2} x_i + \frac{m_i}{t} \frac{d}{dt} x_i + k_{2Oi}(x_i - z) + k_{4Oi}(x_i - z)^3 +$$

$$\sum_{j=1}^{6} k_{2Ij}(2x_i - x_{i+j} - x_{i-j}) + \sum_{j=1}^{6} k_{4Ij}\{(x_i - x_{i+j})^3 + (x_i - x_{i-j})^3\}$$

$$+ \frac{1}{2} e_0 \frac{l_i w}{(d + x_i - z)^2} V^2 = 0 \qquad (1)$$

where for the di-element case $i$ (odd) refers to the long cantilevers and $i$ (even) to the short ones. The displacement at the tip of the $i^{th}$ cantilever end is $x_i$, $m$ is the mass, $t$ is the damping time, $k_{2Oi}$ and $k_{2Ii}$ are the onsite and intersite linear spring constants, $k_{4Oi}$ and $k_{4Ii}$ are the onsite and intersite nonlinear spring constants. The interactions with up to six neighbors are necessary to describe the experimental dispersion curves. To estimate the dispersion curves, standing wave patterns are taken as a function of a driver frequency at low excitation level. The wavenumber of each mode is estimated from the spatial Fourier transformation of the imaged pattern. A cantilever mass is estimated from its volume, and $t$ is calculated from the linear resonance line-width. The last term in Eq. (1) is the electrostatic interaction. Here $e_0$ is the dielectric constant of vacuum, $l$ is the length of the cantilever, $d$ is the gap between the cantilever and the substrate, and $V$ is the voltage applied to the cantilevers. The PZT drives the cantilevers in the array at frequency $f_d$ by modulating their equilibrium positions according to $z = z_0 \cos(2\pi f_d t)$.

By comparing ILM/DB experiments with simulations estimates can be obtained for the onsite nonlinear parameter $k_{4O}$ the nearest neighbor intersite nonlinear parameter $k_{4I1}$. The estimated intersite $k_{4II}$ is about 100 times larger than the estimated onsite $k_{4O}$ consistent with the idea that they are due to stretching of the overhang-membrane by the overhang distortion and cantilever bending, respectively. For an attractive interaction between the ILM/DB and the impurity mode in a soft nonlinear array, in addition to the estimates described above it has also been assumed that $k_{4Ii}/k_{2Ii} = k_{4II}/k_{2II}$ so that the long-range nonlinear coupling has the same range dependence as for the linear coupling. Without these two conditions the large amplitude ILM/DB cantilevers become trapped by the electrostatic potential.

A summary of our simulation findings to characterize the interaction between a laser-induced impurity mode and an ILM/DB stemming from the band states are summarized in Fig. 4. For the harmonic lattice there are only linear band states, which are represented by the rectangle in Fig. 4(a). Also shown are linear impurity modes, which are produced below or above this band when the linear spring constant is decreased or increased, respectively. For the anharmonic lattice the sign of the impurity-ILM/DB interaction depends on the frequency relationship between the ILM/DB, the band states, and the impurity mode. Figure 4(b) displays the results for the hard nonlinear lattice. A stationary ILM/DB is shown. Simulations show that an ILM/DB is attracted (A) to the impurity mode when the impurity mode frequency is between the top of the band and the ILM/DB. If after trapping the ILM/DB at the impurity mode the impurity spring constant is returned slowly ($t_{transition} = 2t$) to the intrinsic value the ILM/DB remains at the trapped site. On the other hand, if the impurity change is rapid the trapped ILM/DB is destroyed. If the impurity frequency is higher than the ILM/DB, it repels (R) the ILM/DB. If the impurity frequency is slightly below the driver frequency and slowly decreased then an ILM/DB appears at the driver frequency and the impurity mode position. This we call seeding. When the impurity mode is below the optic branch it always repels (R) the ILM/DB. Here the speed of change in the impurity mode frequency does not influence the repulsion of the ILM/DB. Thus, these techniques can be used to repel and/or attract ILM/DB.

If the lattice system has a soft nonlinearity, then stationary ILM/DB can be created below the band states as shown in Fig. 4(c). The conditions for the impurity level-ILM/DB repulsion/attraction (R/A) are a vertical mirror image about the center of the band states for the hard nonlinear case just described.

The repulsive/attractive interaction between the focused laser beam and the ILM/DB in the hard/soft nonlinear cantilever array has been observed and explained in terms of an ILM/DB-impurity mode interaction. The local heating by the laser creates the optically-induced real time impurity mode. The shepherding effects with ILM/DB illustrated here for the nonlinear cantilever array should be applicable to any nonlinear discrete lattice, even those at the atomic scale. Particularly noteworthy is the demonstration of optical tweezers applied to localized dynamical energy instead of particles; similar effects may be expected for nanoscale ILM/DB interacting with a scanning force microscope tip at the surface of an atomic lattice.

## ACKNOWLEDGMENTS
We thank Dr. Alan K. Chin at Axel Photonics for providing the semiconductor lasers used in this study. (AJS) acknowledges helpful discussions with Professor J. B. Page. Supported by NSF-DMR and CCMR. The sample fabrication was performed at the Cornell Nanofabrication Facility, also supported by the NSF.

**References**

[1] Sievers A.J. and Takeno S., *Phys. Rev. Lett.*, **61** (1988) 970.
[2] Kivshar Y.S., *Opt. Lett.*, **18** (1993) 1147.
[3] MacKay R.S. and Aubry S., *Nonlinearity*, **7** (1994) 1623.
[4] Sievers A.J. and Page J.B., in *Dynamical Properties of Solids* (Edited by G.K. Horton and A.A. Maradudin), p. 137. North Holland, Amsterdam (1995), Vol. 7.
[5] Floría L.M., Marin J.L., Martínez P.J., Falo F., and Aubry S., *Europhys. Lett.*, **36** (1996) 539.
[6] Flach S. and Willis C.R., *Phys. Repts.*, **295** (1998) 181.
[7] Peyrard M., *Physica D*, **119** (1998) 184.
[8] Trías E., Mazo J.J., and Orlando T.P., *Phys. Rev. Lett.*, **84** (2000) 741.
[9] Binder P., Abraimov D., Ustinov A.V., Flach S., and Zolotaryuk Y., *Phys. Rev. Lett.*, **84** (2000) 745.
[10] Sato M., Hubbard B.E., Sievers A.J., Ilic B., Czaplewski D.A., and Craighead H.G., *Phys. Rev. Lett.*, **90** (2003) 044102.
[11] Mandelik D., Eisenberg H.S., Silberberg Y., Morandotti R., and Aitchison J.S., *Phys. Rev. Lett.*, **90** (2003) 253902.
[12] Fleischer J.W., Carmon T., Segev M., Efremidis N.K., and Christodoulides D.N., *Phys. Rev. Lett.*, **90** (2003) 023902.
[13] Ustinov A.V., *Chaos*, **13** (2003) 716.
[14] Sato M. *et al.*, *Chaos*, **13** (2003) 702.
[15] Fleischer J.W., Segev M., Efremidis N.K., and Christodoulides D.N., *Nature*, **422** (2003) 147.
[16] Swanson B.I. *et al.*, *Phys. Rev. Lett.*, **82** (1999) 3288.
[17] Schwarz U.T., English L.Q., and Sievers A.J., *Phys. Rev. Lett.*, **83** (1999) 223.
[18] Xie A., van der Meer L., Hoff W., and Austin R.H., *Phys. Rev. Lett.*, **84** (2000) 5435.
[19] Sato M., English L.Q., Hubbard B.E., and Sievers A.J., *J. Appl. Phys.*, **91** (2002) 8676.
[20] Younis M.I. and Nayfeh A.H., *Nonlinear Dynamics*, **31** (2003) 91.
[21] Buser R.A. and de Rooij N.F., *Sens. Actuators*, **17** (1989) 145.
[22] Rouxel T., Sangleboeuf J.-C., Huger M., Gault C., Besson J.-L., and Testu S., *Acta Materialia*, **50** (2002) 1669.

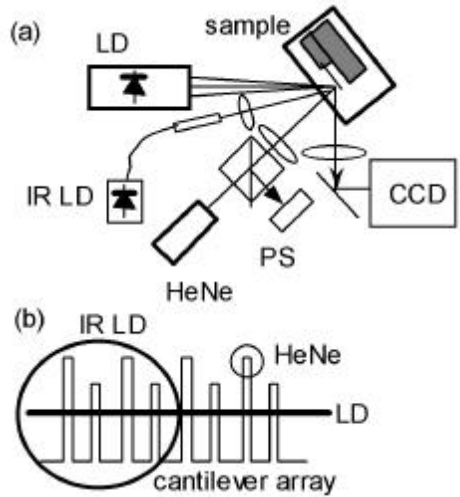

Figure 1. (a) Experimental setup for observing the ILM/DB-impurity mode interaction. The sample in vacuum is driven by a PZT. A 1D CCD camera observes the reflected beam of a laser line illuminator (laser diode=LD). A sharply focused HeNe laser beam can be used to measure the linear vibrational response with the laser position sensor (PS). A fiber pigtailed IR LD is used to locally heat a few elements and produce a movable, gap impurity mode in the cantilever array. (b) Schematic representation of the three kinds of laser beams on a di-element cantilever surface.

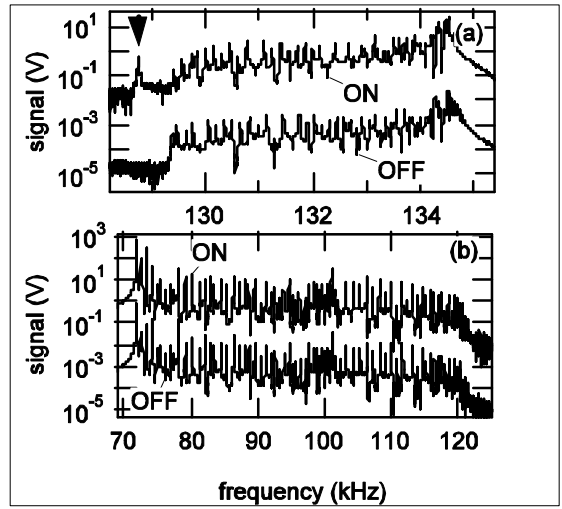

Figure 2. Measured normal modes of a di-element cantilever array with and without local laser heating. The linear response spectra of one cantilever for (a) optic and (b) acoustic type bands are shown. The spectra for the laser heated cases are shifted up for clarity. The impurity mode is created below the upper branch when the sample is locally heated with the IR laser, as indicated by the arrow. The IR laser power is 100 mW.

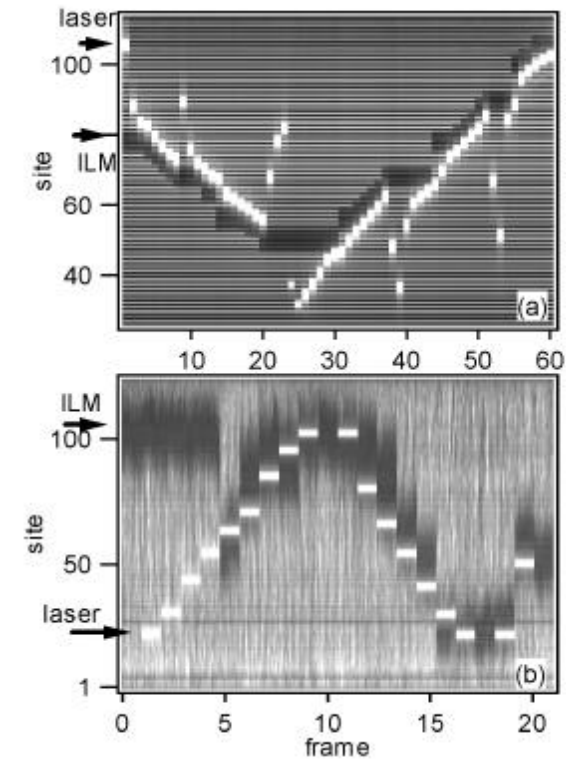

Figure 3. Laser manipulation of an ILM/DB from the band states of a cantilever array with anharmonicity of different signs. The images are constructed from several chronologically ordered still pictures. The dark region in each frame corresponds to a highly excited ILM/DB. The white rectangle identifies the heating laser spot. See arrows. (a) Repulsive interaction in a hard nonlinear di-element cantilever array. As the laser spot approaches, the ILM/DB is repelled and hops away. When the laser spot is far from the ILM/DB, it remains fixed is shown in the frames around 10, 40, and 52. When the laser power is low it can pass through the ILM/DB without interaction (frames around 23). (b) Experimental study of the attractive interaction between the laser-induced impurity mode and the nonlinear ILM/DB in a soft nonlinear mono-element array. The soft nonlinearity is produced by the addition of a dc potential. Due to the large bandwidth of this branch, the ILM/DB oscillates about its equilibrium position. The ILM/DB, it is attracted and captured by the laser spot in frame 5. As the laser spot moves so does the captured ILM/DB. When the laser is turned off, the ILM/DB remains fixed (see frames 10, 18 and 21).

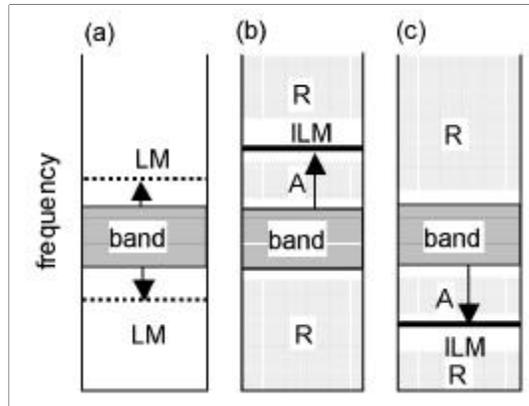

Figure 4. Illustration summarizing the impurity mode-ILM/DB interaction results. (a) Linear case. Impurity modes appear above or below the band states depending on the sign of the linear spring constant change. (b) For the nonlinear array with hard anharmonicity a stationary ILM/DB is created above the band states. When the impurity mode is either below the band states or above the ILM/DB frequency, the ILM/DB is repelled. When the impurity mode is above the band states but below the ILM/DB frequency, the gray region, the ILM/DB is attracted. (c) For the nonlinear array with soft anharmonicity the stationary ILM/DB is created below the band states. The conditions for the repulsion/attraction (R/A) interaction are now inverted with respect to case (b).